\documentclass[%
 reprint,
 amsmath,amssymb,
 aps,
]{revtex4-1}

\usepackage{graphicx}
\usepackage{dcolumn}
\usepackage{bm}
\usepackage{pifont}
\usepackage{subfiles}
\usepackage{units}
\usepackage{hyperref}
\begin{document}

\newcommand*\rfrac[2]{{}^{#1}\!/_{#2}}
\setlength{\textfloatsep}{20pt plus 5.0pt minus 3.0pt}

\preprint{APS/123-QED}

\title{The Angular Power Spectrum of Heavy Ion Collisions}

\author{M. Machado$^{1,2}$}
 \author{P. H. Damgaard$^{1,2}$}%
 \author{J. J. Gaardh\o je$^2$}
 \author{C. Bourjau$^2$}
 \affiliation{%
 	Niels Bohr International Academy$^1$ and Discovery Center$^2$, Niels Bohr Institute, Blegdamsvej 17, DK-2100 Copenhagen, Denmark
}%
 




\date{\today}

\begin{abstract}
Particles produced in heavy ion collisions carry information about anisotropies present already in the early state of the system and play a crucial role in understanding the Quark Gluon Plasma and its evolution. We explore the angular power spectrum of particle multiplicities in such heavy ion collisions to extract fluctuations in particle multiplicities on the surface a sphere. Results are presented for Pb-Pb data at $\sqrt{s_{NN}} = 2.76\unit{TeV}$, extracted from the ALICE open data portal. We find that odd modes of the power spectrum display a power-law behavior with corresponding index $\beta$, which is found to be close to unity. We also demonstrate that the angular power spectrum allows us to extract accurately the flow coefficients of non-central collisions.    
\end{abstract}

\maketitle


\section{\label{sec:level1}Introduction}

	Heavy ion collisions at ultrarelativistic energies, such as currently studied at the top energy ranges of the Large Hadron Collider (LHC) at CERN, can typically generate more than 20,000 charged particles in each central collision \cite{Adam:2016ddh}. In the first instants of the collision a strongly interacting Quark Gluon Plasma (QGP) is formed \cite{Arsene:2004fa,Back:2004je,Adams:2005dq,Adcox:2004mh} which subsequently expands as an isolated system, cools, and hadronizes. A central current theme is the study of the collective expansion of the QGP, which can be described by viscous hydrodynamics and which provides information on the viscosity and other transport properties of this exotic state of matter. A powerful analysis method relies on the understanding of particle production and momentum resultant from colliding ions which hit each other with intermediate impact parameters. The azimuthal distribution of the charged particles created in the overlap (participant) zone can be characterized by flow coefficients $v_n$  of the harmonic modes contributing to the Fourier decomposition of the distribution \cite{Adam:2016izf}.

	In ref. \cite{cmb_heavy_paper1} it was proposed to analyze large-multiplicity heavy ion collisions by means of statistical tools developed for the study of the Cosmic Microwave Background in cosmology. The idea is obvious: to view the QGP as the opaque very early universe, and to see the hadronization as emanating from the ``last scattering surface'' of the QGP. From that stage the color-singlet hadrons can escape and eventually, perhaps after residual fragmentations, reach the particle detector which we parametrize as in the Mollweide map of the sky. In the case of the Cosmic Microwave Background the event is unique and it is naturally analyzed by expanding the observed temperature distributions in terms of spherical harmonics, viewing the distribution as a function of two angular variables (polar angle $\theta$ and azimuthal angle $\phi$). Just as the associated angular power spectrum of the Cosmic Microwave Background contains information on the expansion of the Universe, we may hope glean information about the system of quark and gluon matter, and its expansion and eventual hadronization, from its angular power spectrum. 

	In this Letter we present such a study of the angular power spectrum based on the analysis of public heavy-ion data from Pb-Pb collisions at $\sqrt{s_{NN}} = 2.76\unit{TeV}$ collected with the ALICE experiment and taken from the CERN open data portal~\cite{open_data, cbourjau}. Earlier works in similar directions have been reported in ~\cite{cmb_heavy_paper1, powspec_paper2} and ~\cite{spanish_paper3}. The present study takes a different approach by basing the analysis entirely on the angular power spectrum of the collisions. Special attention needs to be paid to the fact that the heavy ion sky is cut by detector limitations in the direction of the polar caps, and we show how to overcome this by subtraction of the $m=0$ mode. For odd $l$-modes we observe a power-law behavior and we define the corresponding critical index $\beta$, a new and potentially interesting observable in heavy ion collisions. Finally, we compare our determinations of flow coefficients $v_n$ with results yielded from more conventional anisotropic flow analyses, finding good agreement and also some intriguing small discrepancies that suggest that a further focus on these methods may provide new insight into the flow patterns observed.

\section{Map and Spectrum}

	Charged particles produced in ultra relativistic heavy ion collisions at the LHC can be described in terms of their azimuthal angle $\phi$ and polar angle $\theta$ between the particle's 3-momentum and the beam axis. Therefore, multiplicity distributions are expressed by a function $f(\theta, \phi)$, which can be expanded in spherical harmonics,
	
	\begin{equation}
	f(\theta, \phi) = \sum_{l=0}^{l_{max}}\sum_{m=-l}^{m=l}a_{lm}Y_{l}^m(\theta, \phi),
	\label{eq:expansion_sph}
	\end{equation} 
	
	\noindent with 
	
	\begin{equation}
	Y_{l}^m(\theta, \phi) = \sqrt{\frac{2l + 1}{4\pi}\frac{(l - m)!}{(l + m)!}}P_{l}^m(\cos(\theta))e^{im\phi},
	\label{eq:sphs}
	\end{equation}
	
	\noindent where $\theta \in [0, \pi]$, $\phi \in [0, 2\pi)$, and $P_{l}^m(\cos(\theta))$ are the associated Legendre Polynomials. The $l$-modes are labeled as multipole moments, with $l = 0$ being the monopole, $l = 1$ the dipole, and so on. Here it is assumed that modes with $l > l_{max}$ hold insignificant signal power.
	
\begin{figure}[!ht]
		\includegraphics[width=0.4\textwidth]{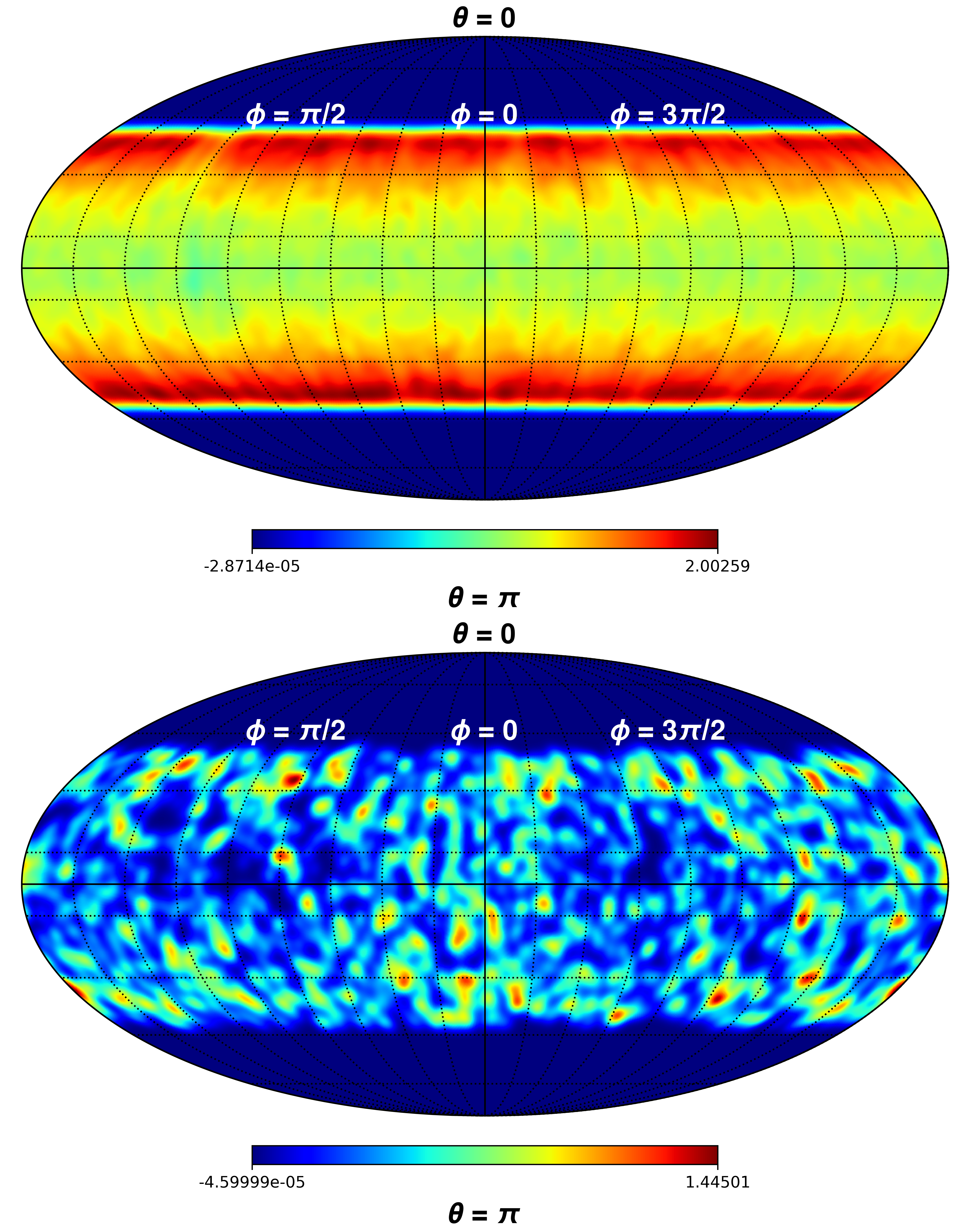}
		\caption{\label{fig:map_singleevt} Maps of all available events (top) and a single heavy ion event divided by it, $f(\theta, \phi)/F^{all}(\theta, \phi)$, (bottom) for 0-5\% centrality at energy $\sqrt{s_{NN}} = 2.76 \unit{TeV}$ within $|\eta| < 0.9$.}
	\end{figure}

	The software HEALPix (Hierarchical Equal Area isoLatitude Pixelation)~\cite{healpix} partitions a spherical surface in pixels of equal area, providing a map of an event's final particle distribution as a pixelation of Eq.~(\ref{eq:expansion_sph}). The number of divisions on the sphere depends on the chosen resolution. Each particle with coordinates ($\theta_j, \phi_j$) maps to a pixel $j$ on the sphere, as shown in Fig.~\ref{fig:map_singleevt}. The color coding follows particle density times map resolution (number of pixels). 
	
	The polar angle relates to pseudorapidity $\eta$ via $\theta = 2\arctan(\exp(-\eta))$. The limitation in pseudorapidity to $|\eta| < 0.9$ is imposed on all charged particles reconstructed with the Time Projection Chamber (TPC) of the ALICE detector ~\cite{why_eta_less_08}. At the 0-5\% centrality, the multiplicity of each event ranges from $\sim2000$ to $\sim3000$ and a little experimentation leads us to choose $l_{max} = 23$. Through HEALPix, the sphere is then divided into 768 pixels of equal area and the region within $|\eta|< 0.9$ ($44^{\mathrm{o}} \lesssim \theta \lesssim 136^{\mathrm{o}}$) for which our data are limited contains 544 pixels. To better visualize where the particles cluster, the maps in Fig.~\ref{fig:map_singleevt} have been smoothed by $ 5^{\mathrm{o}} $: particles were bundled in gaussian beams with full width at half maximum of $\sim0.09$ radians ($5^{\mathrm{o}}$).
	
	In order to correct for detector efficiency, we have chosen the following approach: first, all events corresponding to a certain centrality were added and thus mapped onto the same Mollweide projection (top of Fig.~\ref{fig:map_singleevt}) regardless of the orientation of their reaction plane. This map is referred to as $F^{all}(\theta, \phi)$. Due to the random azimuthal orientation of each collision's reaction plane, all remaining anisotropies along $\phi$ in $F^{all}(\theta, \phi)$ may then be attributed to detector defects. Second, each event map was divided by the all-event map within the $|\eta| < 0.9$ range. Thus, given that $f(\theta, \phi)$ represents an event map, $f(\theta, \phi) \to \bar{f}(\theta, \phi) = f(\theta, \phi)/F^{all}(\theta, \phi)$, with $\bar{f}(\theta, \phi)$ the corrected map. The latter is depicted in Fig.~\ref{fig:map_singleevt} (bottom) and we use it to determine the coefficients $a_{lm}$:
	
	\begin{equation}
	a_{lm} = \frac{4\pi}{N_{pix}}\sum_{j = 0}^{N_{pix} - 1}Y^*_{lm}(\theta_j, \phi_j)\bar{f}(\theta_j, \phi_j),
	\label{eq:alms}
	\end{equation} 
	
	\noindent where $N_{pix}$ represents the total number of pixels. Additionally, it should be remarked that this is the \textit{zeroth order estimator}, as pixelating $f(\theta, \phi)$ ($f(\theta, \phi) \to f(\theta_j, \phi_j)$) corresponds to taking its average within each pixel with surface area $\Omega_{pix}$~\cite{healpix}. The higher order estimators are implemented in the facilities of HEALPix. From Eq.~(\ref{eq:alms}) the angular power spectrum $C_l$ of an individual heavy ion event is then defined by
	
	\begin{equation}
	C_l = \frac{1}{2l + 1}\sum_{m = - l}^{m = l}|a_{lm}|^2.
	\label{eq:cls}
	\end{equation}
	
	\noindent The multipole moments relate to the angular scale $\alpha$ of the distribution through $\alpha = \frac{180^{\mathrm{o}}}{l}$.	
	\begin{figure}[!ht]
		\includegraphics[width=0.4\textwidth]{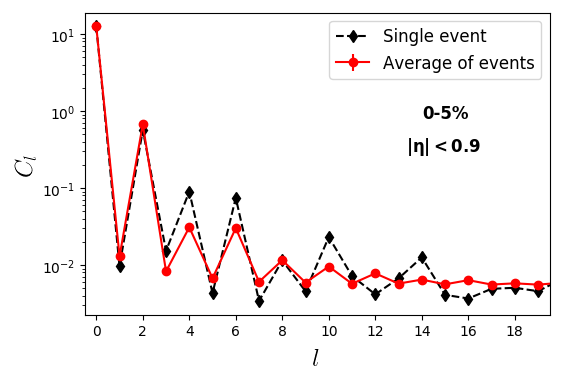}
		\caption{\label{fig:powspecs05} Angular power spectra of the event in Fig.~\ref{fig:map_singleevt} (diamonds) and of a 8000-event average (circles).}
	\end{figure}
	
	The fixed value $C_0 = 4\pi$ for all events is a consequence of the chosen normalization, obtained by dividing $\bar{f}(\theta,\phi)$ by the event multiplicity and multiplying it by $N_{pix}$.  In the present case, $\bar{f}(\theta, \phi)$ is a piecewise function: while well defined within $44^{\mathrm{o}} \lesssim \theta \lesssim 136^{\mathrm{o}}$, it takes a null value otherwise.  This causes a suppression of $l = 4n$, $n = 1, 2,\ldots$ relative to the other even modes, as seen in Fig.~\ref{fig:powspecs05}, an effect first identified in ref. \cite{spanish_paper3}. 
	
\begin{figure}
		\includegraphics[width=0.4\textwidth]{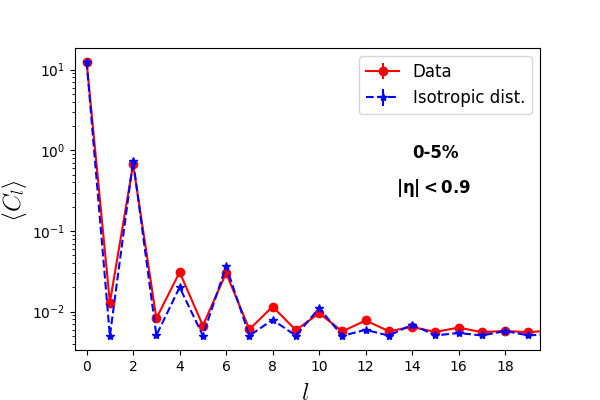}
		\caption{\label{fig:iso_vs_data-full} Comparison between averaged power spectra of heavy ion data and isotropic distributions for 0-5\% centrality.}
	\end{figure}

\begin{figure}[!ht]
		\includegraphics[width=0.4\textwidth]{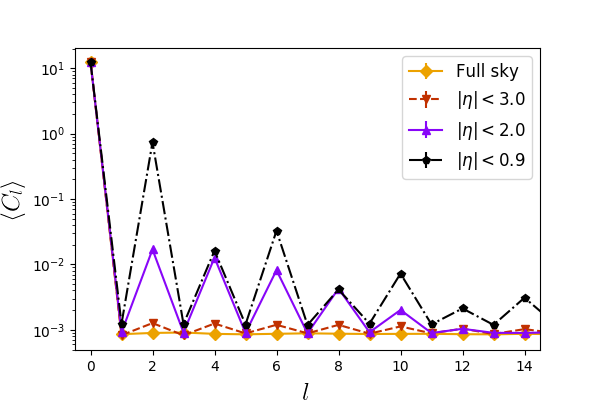}
		\caption{\label{fig:comp_iso_cuts} Comparison between averaged power spectra of isotropic distributions for different $\eta$ range values.}
	\end{figure}

	In order to test these important edge effects on the power spectrum, we have generated 8000 distributions that are isotropic on the surface of a unit sphere and have the same multiplicities as the 0-5\% centrality. We then computed $C_l$ for each event after submitting $(\theta_j, \phi_j)$, $j = 0,...,N_{pix}$ to the cut $44^{\mathrm{o}} \lesssim \theta \lesssim 136^{\mathrm{o}}$. The resulting averaged power spectrum is shown in Fig.~\ref{fig:iso_vs_data-full} and compared to actual data. For a sphere with smooth uniform distribution $Y_{00} = 1/\sqrt{4\pi}$ holds all the power. Consequently, $C_0 = 4\pi$ and $C_l = 0$ for $l > 0$. This remains approximately true for a discrete distribution that is isotropic on average, which we label \lq full sky\rq: the monopole yields a power spectrum value of $4\pi$, while the other moments have $C_l \sim 10^{-3}$ (see Fig.~\ref{fig:comp_iso_cuts}), only differing from zero due to the chosen finite multiplicity - the $\langle C_l \rangle$ in Fig.~\ref{fig:comp_iso_cuts} come from events with multiplicity of $\sim 15000$. As the latter increases, $C_l$ approaches zero for $l > 0$; this issue shall be addressed in more detail in ref. \cite{Meera}.   
	
	It is striking that the isotropic distributions shown in Fig.~\ref{fig:iso_vs_data-full}, which have the same multiplicities as the real events from 0-5\% centrality, have almost identical averaged power spectra. The modes $l = 4n$, $n = 1, 2,...$ are strongly suppressed relative to the other even $l$, demonstrating that such feature originates from the limited detector acceptance. Tiny differences can be observed: $C_{4,8,12}$ have slightly higher values for data than the isotropic distributions. This suggests that there are more fluctuations within a solid angle $\Omega = \pi/2$ for real data than for the isotropic case. 
	
	Since gross features of the angular power spectrum are so well reproduced by simulations, it is straightforward to conclude that the suppression of modes with $l = 4n$ is an artifact of data being limited to the $\theta$-range shown. This is demonstrated in Fig.~\ref{fig:comp_iso_cuts}, where we show averaged power spectra of isotropic distributions with same multiplicity for different cuts in $\eta$. As the range in $\eta$ narrows, even $l$-modes become more enhanced.

\begin{figure}[!ht]
		\includegraphics[width=0.4\textwidth]{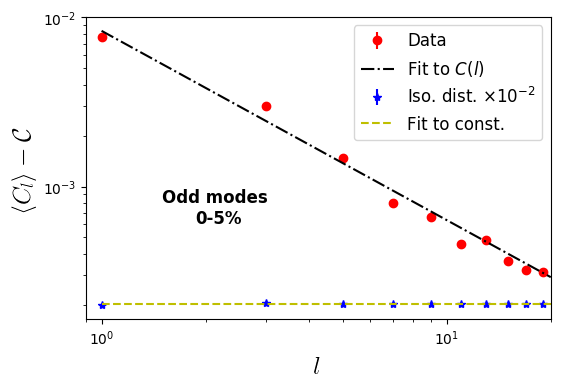}
		\caption{\label{fig:iso_vs_data-odd} Comparison between averaged odd modes of heavy ion data and isotropic power spectra for 0-5\% centrality; data is fit to $C(l)$.}
	\end{figure}

\begin{figure}[!ht]
		\includegraphics[width=0.4\textwidth]{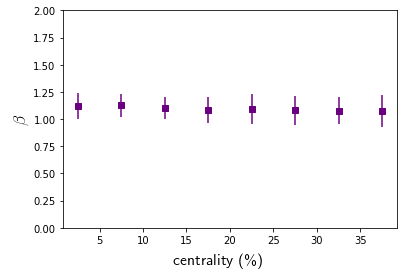}
		\caption{\label{fig:params_centr} Parameter $\beta$ resultant of fit to $C(l)$ as a function of centrality.}
	\end{figure}

	For isotropic distributions, $\langle C_l\rangle$ for odd $l$-values are consistently low, as expected on account of parity symmetry and as seen in Fig.~\ref{fig:comp_iso_cuts}. Real data, even those associated with central collisions, show a markedly different behavior as shown in Fig.~\ref{fig:iso_vs_data-odd} for the case of 0-5\% centrality. Because the data appear to follow a power law we have performed fits of averaged power spectra to the function $C_l = A\cdot l^{-\beta} + \mathcal{C}$. The exponent $\beta$ appears to be independent of centrality and it is very close to unity (a best fit to a constant yields a value $~1.068$, consistent with unity within $1\sigma$) as shown in Fig.~\ref{fig:params_centr}. The standard deviation of each $\beta$ was calculated from the variance of the parameter estimate.  
	
\section{Power Spectrum for $m \neq 0$}

	The limitation imposed by the acceptance of the TPC leads us to explore a new strategy that corrects the data from artificial suppressions of those even-$l$ modes that are just a consequence of the geometric limitation. Since we are free to define the angular power spectrum we need not tie ourselves to the standard definition employed in situations where there is an essentially uniform map of the full sky, with smaller fluctuations on top. We therefore define a modified angular power spectrum where the 'global' $m=0$ mode is excluded:
	
	\begin{equation}
	C^{m\neq0}_l = \frac{1}{2l+1} \sum_{m=-l}^{m=l} |a_{lm}|^2 - \frac{|a_{l0}|^2}{2l+1}.
	\label{eq:cl_mdz} 
	\end{equation}
	
\begin{figure}[!ht]
		\includegraphics[width=0.4\textwidth]{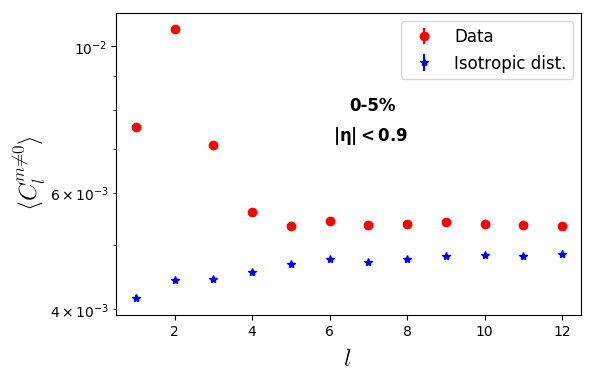}
		\caption{\label{fig:iso-vs-data_mdz} Averaged power spectra for $m\neq0$ at 0-5\% centrality. Real data is represented by the circles, while isotropic distributions are the stars.}
	\end{figure}

\begin{figure}[!ht]
		\includegraphics[width=0.4\textwidth]{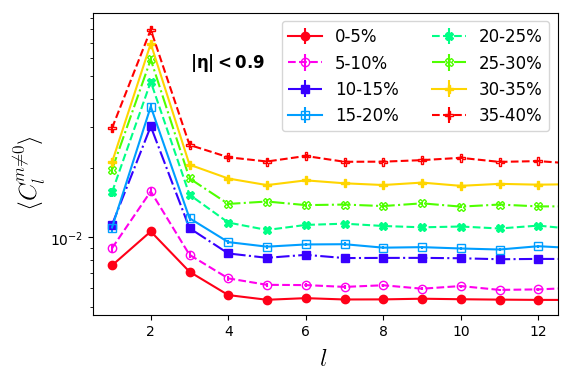}
		\caption{\label{fig:avgall_mdz} Averaged power spectra for $m\neq0$ at all centralities.}
	\end{figure}

	This definition is designed to remove detector acceptance effects while keeping essential physical information of all modes except those corresponding to $m=0$. Indeed, from Fig.~\ref{fig:iso-vs-data_mdz} we see that the averaged power spectrum for the isotropic distributions is close to trivial, as expected. On simulated data there are no microscopic physical mechanisms to introduce a non-trivial signal and the power spectrum is essentially flat. On the other hand, real data display a clear peak in $l=2$, aside from also the enhanced $\langle C^{m\neq0}_1 \rangle$ and $\langle C^{m\neq0}_3 \rangle$ values. The same pattern occurs for all centralities (Fig.~\ref{fig:avgall_mdz}), suggesting the presence of real anisotropies in the considered events. The continuous enhancement in magnitude as the spectra become associated with more peripheral collisions is mainly due to decreasing multiplicities, since map sparsity leads to an increase in fluctuations, thus enhancing $C_l$ values.	
	
\section{Flow Extraction}

Matter produced in heavy ion collisions exhibits strikingly what is known as collective flow~\cite{PhysRevLett.52.1590, PhysRevLett.80.4136}, indicating that anisotropies present in the early stages of the collision result in increased particle emission in certain directions. In the presence of transverse flow (perpendicular to the collision axis), the final particle distribution along the azimuthal direction can be decomposed as~\cite{Voloshin:1994mz, Poskanzer:1998yz}:

\begin{equation}
\frac{dN}{d\phi} \propto \frac{1}{2\pi}\left[ 1 + 2\sum_{n=1}^{\infty} v_n \cos(n(\phi-\psi_n)) \right],
\label{eq:flow_dist}
\end{equation}

\noindent where $v_n$ are the flow coefficients and $\psi_n$ the symmetry planes corresponding to the different $n$ modes. Coefficients $v_1$, $v_2$ and $v_3$ are denoted \textit{directed}, \textit{elliptic} and \textit{triangular} flow, respectively.

Given our observed enhancement of low-$l$ modes in the averaged power spectra $\langle C^{m\neq0}_l \rangle$ as compared to the isotropic case (see Fig.~\ref{fig:iso-vs-data_mdz}), it is natural to ask if this could be caused by flow and thus potentially provide an alternative method for determining it. To investigate this question, we associate $\bar{f}(\theta, \phi)$ with the left hand side of Eq.~\ref{eq:flow_dist} for $44^{\mathrm{o}} \lesssim \theta \lesssim 136^{\mathrm{o}}$ ($|\eta| < 0.9$) and zero otherwise, while constant in $\theta$. From the expansion in spherical harmonics (Eq.~\ref{eq:expansion_sph}) and the analytical form of Eq.~\ref{eq:alms} we find

\begin{align}
a_{l0} &= b_{l0} &\text{ for } m = 0\nonumber,\\
a_{lm} &= b_{lm}\cdot v_{|m|} e^{-im\psi_{|m|}} &\text{ for } m \neq 0,
\label{eq:fourier_alms}
\end{align}

\noindent where

\begin{equation}
b_{lm} = \sqrt{\frac{2l + 1}{4\pi}\frac{(l-m)!}{(l+m)!}}\int_{\theta_i}^{\theta_f}\sin{\theta}P_{lm}(\cos{\theta})d\theta
\label{eq:blm_coefs}
\end{equation}

\noindent for all $l$ and $m$. Here $\theta_i$, $\theta_f$ define the initial and final values of the interval where $\bar{f}(\theta, \phi)$ is nonvanishing. Given that $\sin{\theta}\cdot P_{lm}(\cos{\theta})$ is symmetric around $\pi/2$, the only surviving coefficients will have $l$ and $m$ sharing the same parity. Additionally, flow coefficients $v_n$ as well as their angles of symmetry planes $\psi_n$ do not contribute to any of the $m=0$ modes. The event plane angle $\psi_n$ does not affect the angular power spectrum since $a_{lm}\propto e^{-in\psi_n}$ and only $|a_{lm}|^2$ enters there.

Solving for $|v_n|$ $n=1,2$, after combining Eq.(\ref{eq:cl_mdz}) with Eqs.(\ref{eq:fourier_alms}, \ref{eq:blm_coefs}), we find

\begin{equation}
|v_n|^2 = \frac{(2n + 1)}{2}\cdot\frac{C_n^{m\neq0}}{|b_{nn}|^2}\cdot\frac{|b_{00}|^2}{4\pi}. 
\label{eq:v23} 
\end{equation} 

\noindent For the case at hand, $|b_{31}|^2$ is of $\mathcal{O}(10^{2})$ smaller than $|b_{31}|^2$, leading to the use of Eq.(\ref{eq:v23}) in the calculation of $v_3$.

Increased sparsity of particles leads to higher $C_l$ values due to larger fluctuations and it is crucial to take this into account when studying their magnitude. We propose to correct for the trivial increase due to limited multiplicity by subtracting the spectrum of isotropic distributions with multiplicities corresponding to the chosen centralities: $\langle C^{m\neq0}_n\rangle \to \langle C^{m\neq0}_n\rangle - \langle C^{m\neq0[iso]}_n\rangle$ in Eq.~(\ref{eq:v23}). To compare, we have generated $10^6$ isotropic Mollweide maps limited to $|\eta| < 0.9$ with corresponding multiplicities. As in all previous cases, the angular power spectrum has been extracted from each map and averaged over all events. 

\begin{figure}[ht]
	\includegraphics[width=0.4\textwidth]{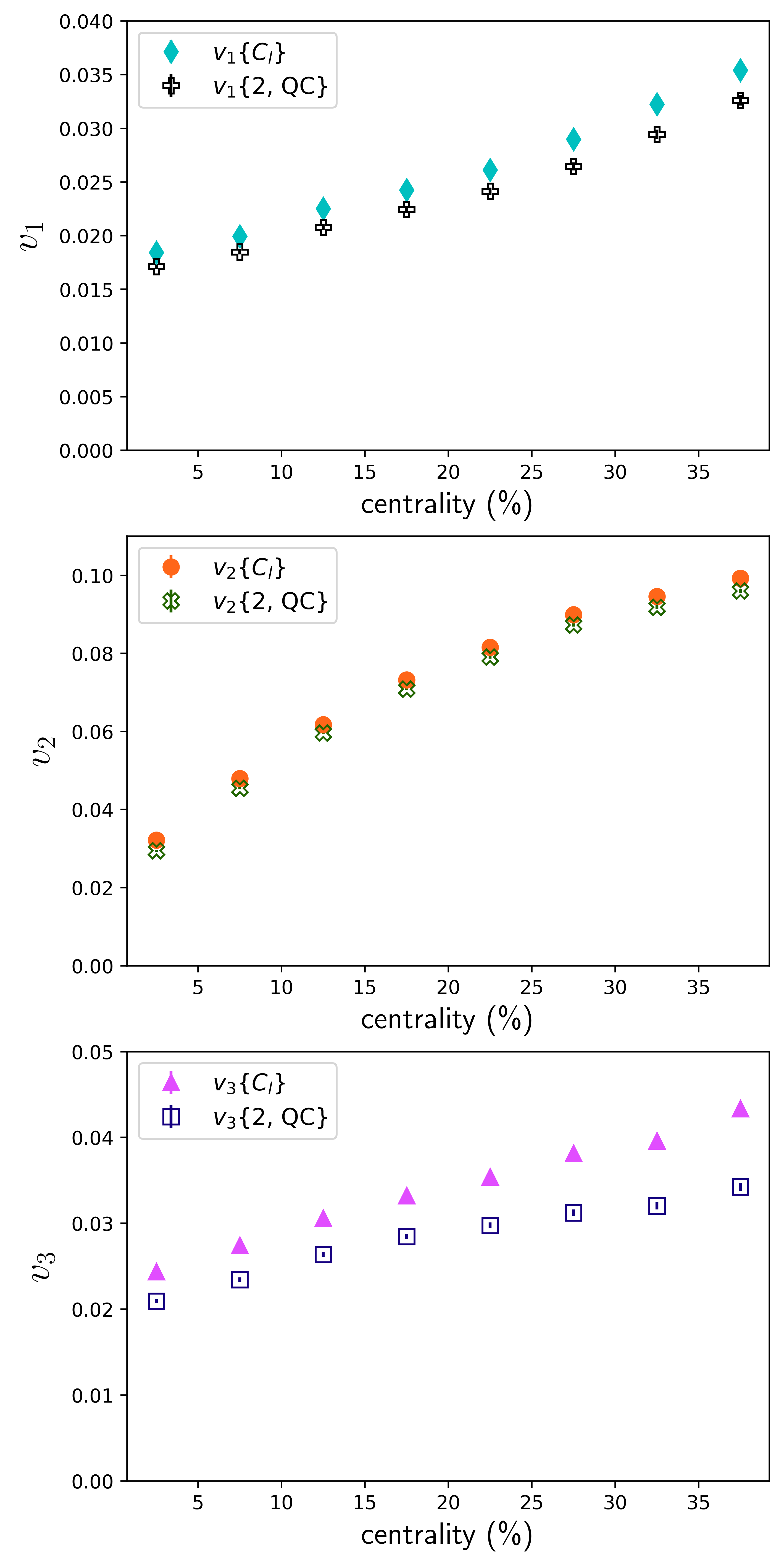}
	\caption{\label{fig:comp_data_vns} Comparison between $v_1$ \textbf{(a)}, $v_2$ \textbf{(b)} and $v_3$ \textbf{(c)} calculated with the power spectrum approach and the Q-cumulants on real ALICE data~\cite{open_data}.}  
\end{figure}

The extraction of leading flow coefficients $v_1$, $v_2$ and $v_3$ from the power spectrum then proceeds as follows: For each centrality we compute $\langle C^{m\neq0}_n\rangle - \langle C^{m\neq0[iso]}_n\rangle$ and substitute it into  Eq.~(\ref{eq:v23}). We have successfully tested the high accuracy of this method with Monte Carlo data that took specific $v_i$'s as input. Using this method we have calculated $v_n$ for the ALICE open data~\cite{open_data} and compared with the \textit{Q-cumulants} method of flow analysis for two-particle correlations~\cite{qcumulants} applied to the same data set. In Fig.~\ref{fig:comp_data_vns} these coefficients are labelled $v_n\{C_l\}$ and $v_n\{2, QC\}$, respectively. For directed and elliptic flows ($v_1$ and $v_2$), numbers agree better than for triangular flow coefficients $v_3$, which show small deviations, especially for peripheral centralities. We stress that when both methods are applied to Monte Carlo simulations, i.e., purely flow scenarios, they agree uniformly. 

The angular power spectrum is simply the two-point correlation function in Fourier space. In that case, computing $C_l$ for events maps like in Fig.~\ref{fig:map_singleevt} (bottom) is akin to correlating pixel windows. The Q-cumulants method employed in this study takes the two-particle correlation function in the azimuthal direction only. Therefore, the first main difference between the two approaches is \textit{what} they correlate: for the former, it is $(\theta_i,\phi_i)$ windows, while for the latter it is $\phi_i$. The Monte Carlo simulations considered $f(\theta, \phi)$ as both factorizable and having the same azimuthal distribution for all $\theta$, so it is not surprising that $v_n\{C_l\}$ and $v_n\{2, QC\}$ would agree. In light of this fact, a second set of MC events was generated with $v_n$ coefficients varying slightly with $\theta$ and the result persisted.

In regards to correlations arising from jets and resonance decays, it escapes the range of this paper to answer how the angular power spectrum responds to such effects. Having said that, it must be emphasized that the scale structure of a jet, for instance, is quite small in comparison to elliptic eccentricity. Therefore, jets are expected to mainly influence higher $l$-modes and have negligible to no contribution to lower $l$. It should also be mentioned that the Q-cumulants calculation in this study did not take into consideration pseudorapidity gaps. That means non-flow contributions have not been suppressed when computing $v_n\{2, QC\}$ either. They may have different responses to non-flow effects, as their two-point functions are computed in distinct dimensions, though that is an issue for future research~\cite{Meera}.


\section{Discussion}

	We have explored some of the powerful methods of Comic Microwave Background analyses when applied to the study of heavy ion collisions with very large particle multiplicities. We have shown that it is crucial to take into account the limitations of detector coverage as compared with the (almost) full-sky coverage of the Cosmic Microwave Background. Detector limitations introduce artificial structures in the angular power spectrum $C_l$ that can swamp the physical information. This holds in particular for multipoles of even $l$ with $m=0$. For the ALICE detector and the publicly available data used in the present study the coverage is roughly between 45$^{\circ}$ and 135$^{\circ}$, leading to a suppression of $l=4n$, $n=1,2,\cdots$ relative to the otherwise enhanced even $l$'s. Other ranges of cuts in the polar angle with otherwise isotropic Monte Carlo distributions demonstrate clearly the relative suppressions of the (shifting) even $l$-modes due simply to the geometry of the detector limitations. The odd $l$-modes, although suppressed on average as compared to the even $l$-modes due to the approximate parity symmetry between multiplicities in the forward and backward directions, provide intriguing new information about the events. Here, there are very pronounced differences between the angular power spectrum of real data as compared to simple distributions based on approximate isotropy (which have essentially vanishing odd components on account of parity). Actual data seem to obey quite accurately a power-law behavior $C_l = A\cdot l^{-\beta} + \mathcal{C}$ over a wide range of odd $l$-values. The exponent $\beta$ appears to be constant, independent of centralities. What could be the origin of such scaling law, restricted to the odd-$l$ sector? When looking in closer details we find that this scaling appears to be directly triggered by the spread in interaction points of the two colliding heavy ions. Although a miniscule effect at the detector level, collisions that occur slightly shifted with respect to the center of the detector lead to a small forward-backward asymmetry in the angular coverage of the event. This issue will be addressed in more details in a forthcoming paper \cite{Meera}. 

	Finally, we have demonstrated that the angular power spectrum can be used to compute flow coefficients $v_n$. Again care must be taken in order to compensate for the limited range of the TPC detector. We have noted that such effects are encoded strongly in the $a_{l0}$ coefficients and an efficient way to eliminate detector limitations in the $\theta$-direction is to compute the angular power spectrum $C^{m\neq0}_l$ without contributions from $a_{l0}$. Flow coefficients were then extracted using the power spectrum for $m\neq 0$ and compared to the cumulant method~\cite{qcumulants} for two-particle correlations. 

	These results show that analyzing heavy ion collisions by means of the angular power spectrum is a promising new avenue. As a measure of two-point correlations between $(\theta_i, \phi_i)$ windows, it could show that particle distributions have $\eta$ and $\phi$ event-by-event dependencies not seen when taking only azimuthal two-point correlations, even with $\eta$ gaps. We have here focused on the simplest of all observables, particle multiplicity. It would be most interesting to extend this analysis to explore how jets can impact the angular power spectrum or how it looks like when considering particles within different $p_T$ intervals.\\ 

\begin{acknowledgments}
	M.M. is most grateful to Ante Bilandzic for crucial help in the first part of this project. We thank Pavel Naselsky, Hao Liu and You Zhou for illuminating discussions. This work was supported in part by the Danish National Research Foundation (DNRF). The research of M.M. is supported by the Conselho Nacional de Desenvolvimento Cient\'{i}fico e Tecnol\'{o}gico (CNPq).
\end{acknowledgments}

\nocite{*}

\bibliography{paper1_powspec_mod}

\end{document}